# Probing oppositely charged surfactant and copolymer interactions by isothermal titration microcalorimetry

**J. Courtois and J.-F. Berret\***

Matière et Systèmes Complexes, UMR 7057 CNRS Université Denis Diderot Paris-VII, Bâtiment Condorcet, 10 rue Alice Domon et Léonie Duquet, 75205 Paris (France)
*Corresponding author jean-francois.berret@univ-paris-diderot.fr

**Abstract.** The complexation between charged-neutral block copolymers and oppositely charged surfactants was investigated by light scattering experiments and by isothermal titration calorimetry (ITC). The copolymer was poly(sodium acrylate)-*b*-poly(acrylamide) and the surfactant dodecyltrimethylammonium bromide (DTAB). In a previous report, we had shown that the copolymers and the surfactants co-assembled spontaneously into colloidal complexes. Depending of the charge ratio $Z = [DTA^+]/[COO^-]$, the complexes were either single surfactant micelles decorated by copolymers, or core-shell hierarchical structures. ITC was performed in order to investigate the thermodynamics of the complex formation. Titrations of copolymers by surfactants and of surfactants by copolymers revealed that the electrostatic co-assembly was an endothermic reaction, suggesting a process dominated by the entropy of the counterions. Here we found that the thermodynamic quantities associated with the reaction depended on the mixing order. When surfactants were added stepwise to copolymers, the titration was associated with the formation of single micelles decorated by a unique polymer. Above a critical charge ratio, the micelles rearranged themselves into 100 nm colloidal complexes in a collective process which displayed the following features : *i)* the process was very slow as compared to the timescale of Brownian diffusion, *ii)* the thermodynamic signature was a endothermic peak and *iii)* the stoichiometry between the positive and negative charges was modified from n = 0.48 (single micelles) to 0.75 (core-shell complexes). When copolymers were added stepwise to surfactants, the titration resulted in the formation of the core-shell aggregates only. In both experiments, the amount of polyelectrolytes needed for the complex formation exceeded the number required to compensate the net micellar charge, confirming the evidence of overcharging in the complex formation.

# 1 - Introduction

Electrostatic complexation between oppositely charged species has attracted considerable attention in recent years because of the possibility to design nano-objects with novel functionalities. One of the pioneering work in this fast-growing field was due to Kataoka and coworkers who disclosed the formation of polyion complex micelles from a pair of oppositely-charged block polypeptides with poly(ethylene glycol) segments [1]. Since this first report, charged-neutral diblock copolymers, also dubbed double hydrophilic copolymers have become the regular components of complexation co-assembly schemes [2-4]. These hydrosoluble macromolecules were found to associate spontaneously with oppositely charged systems, such as surfactants [5-7], polymers [8,9], counterions [10] and nanoparticles [11,12], yielding "supermicellar" aggregates with core-shell structures. Because this approach of nanofabrication covers a large spectrum of sizes, hybrid structures based on electrostatic co-assembly appeared as promising for applications dealing with drug delivery, functionalization, catalysis, biomedicine, sensors and membranes.

As far as electrostatic co-assembly using block copolymers is concerned, most experiments focused so far on the microstructure of the mixed aggregates. This microstructure was studied using electron microscopy and small-angle light, neutron and x-ray scattering spectroscopy [4]. In this respect, static and dynamic light scattering experiments have become a prevailing technique for the measurement of the molecular weight, aggregation number and hydrodynamic diameter of such complexes. In contrast, very little is known about the thermodynamics of association between oppositely charged species. Thermodynamic surveys of electrostatic complexation were generally performed using Isothermal Titration Calorimetry (ITC), a technique which permits the determination of the binding enthalpy and binding constant of the reaction between a macromolecule and a ligand [13,14]. DNA and oligonucleotides complexed with oppositely surfactants were extensively studied by ITC because of their potential applications as non-viral gene delivery vehicles [15-18]. Concerning oppositely charged surfactants and polymers [19-22], and in particular the system composed of poly(sodium acrylate) (PANa) and dodecyltrimethylammonium bromide (DTAB), Wang and coworkers have reported a series of calorimetric results that demonstrate the endothermic character of the complexation [23,24]. Similar results were found with poly(acrylic acid)-*b*-poly(ethylene oxide) copolymers complexed with calcium ions [10]. For sake of completeness, it should be mentioned that ITC experiments were also performed on polyelectrolyte pairs entering in the fabrication of alternating multilayers [25,26], as well as on counterions [27,28], nanoparticles [16,29,30] and proteins [31,32] complexed with ion-containing macromolecules.

Some years ago, we studied the interactions between poly(sodium acrylate)-*b*-poly(acrylamide) block copolymers (PANa-*b*-PAM) and the oppositely charged dodecyltrimethylammonium bromide (DTAB). For this system, a combination of the small-angle neutron and light scattering techniques allowed us to identify the core-shell microstructure of DTAB/PANa-*b*-PAM aggregates [6,33,34]. The core of the aggregates was described as a dense coacervate microphase comprising the polyelectrolyte blocks and surfactant micelles. Within the core, the intermicellar distance was determined to be equal to the micellar diameter (4 nm), suggesting a compact arrangement of the spheres. The corona was identified as being made from the neutral blocks. A thorough analysis of neutron data revealed that the amount of polyelectrolytes needed to build the hierarchical colloids always exceeded the number that was necessary to compensate the charge of the micelles. This result was interpreted as an evidence of overcharging in the complexation process [35].

In the present paper, we re-examine the DTAB/PANa-*b*-PAM system using isothermal titration calorimetry. Two types of experiments were performed with ITC, the titration of the copolymers by the surfactants and the reverse. Both experiments exhibit classical sigmoidal decreases of the exchanged heat as a function of the added ligands. These sigmoids were found to be well accounted for by the Multiple Non-interacting Sites model [13, 36]. In this paper, we confirm that the complexation between poly(acrylic acid) segments and dodecyltrimethylammonium bromide is endothermic [23, 24]. We also found that the mixing order is important for the electrostatically driven structures. By a combination of light scattering and ITC, we show that the titration of a block copolymer solution by surfactants results in a two-step process. With increasing DTAB, there is first the formation of single surfactant micelles complexed by one copolymer, and then these micelles rearrange themselves into 100 nm-colloidal core-shell complexes. The signature of the onset of clustering are very slow thermal relaxations and the occurrence of a secondary endothermic peak. We show that the reaction stoichiometry for the micelles and for the complexes are different, 0.48 ± 0.07 versus 0.75 ± 0.05, and confirm the overcharging hypothesis anticipated from structural studies. The titration of a surfactant solution by copolymers gives rise to a unique process associated with the spontaneous growth of hierarchical aggregates.

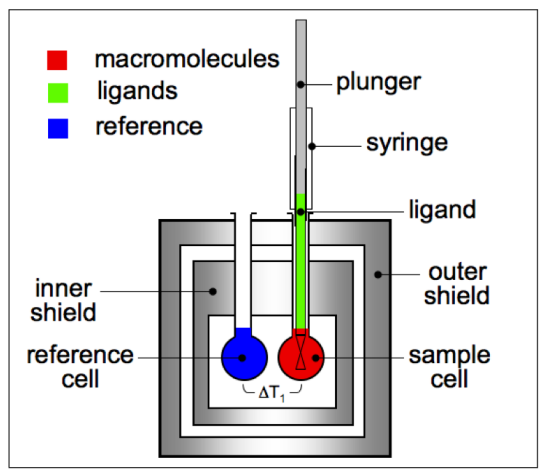

***Figure 1 :*** *Schematic diagram of an ITC instrument* [13, 14]. *As ligands are injected by the syringe to the adiabatic cell, heat can be either released or taken up depending on the interaction process with the macromolecules. A feedback circuit modifies the power applied on the measuring cell so as the difference with the reference remains constant. The heat supplied to the sample cell is the signal measured in ITC experiment. An example of this signal is shown in Figure 2a.*

## 2 - Theoretical Background

In a typical ITC experiment, a solution of a ligand is titrated into a solution of a binding macromolecule. The concentrations of ligands and macromolecules are noted [L] and [M] respectively, $V_0$ being the volume of the cell containing the macromolecules. Here, we present the predictions of the Multiple Non-interacting Sites (MNIS) model [13, 36]. This model assumes that the macromolecules are comprising several anchoring sites that ligands can bind in an uncorrelated way. The probability of binding for a ligand is assumed to be independent on the rate of occupation of the other site of the same macromolecule. The physico-chemical reaction $M + L \to ML$ between macromolecules and ligands is associated with an absorption or a release of heat that is proportional to the amount of binding. The reaction $M + L \to ML$ is characterized by a binding constant $K_b$ and by a reaction stoichiometry noted n. n denotes the number of non-interacting binding sites available on each macromolecule. n = 1 defines the single site binding model [13]. Because in an ITC assay the heat exchange is measured incrementally by successive addition of small amounts of ligands (of the order of 10 µl of ligand solution), the measured quantity represents the derivative of the heat with respect to the number of mole of ligands injected, noted $dQ/dn_L$ where $n_L = V_0 \times [L]$. In the framework of the MNIS model, $dQ/dn_L$ reads [13, 36]:

$$\frac{dQ}{dn_L} = \frac{1}{2}\Delta H_b \left(1 + \left(1 - \frac{[L]}{n[M]} - \frac{1}{nK_b[M]}\right)\left(\left(1 + \frac{[L]}{n[M]} + \frac{1}{nK_b[M]}\right)^2 - \frac{4[L]}{n[M]}\right)^{-1/2}\right)$$

(1)

where $\Delta H_b$ denotes the enthalpy of binding. Eq. 1 is obtained by assuming that the heat released or adsorbed during the titration is proportional to the fraction of bound ligands. For electrostatic complexation, it is convenient to introduce the charge ratio Z between the opposite charges present in the measuring cell. If the concentrations [L] and [M] in Eq. 1 are expressed as the concentrations of charged ligands and charged monomers respectively, one gets $Z = [L]/[M]$. According to these assumptions and posing $r = 1/K_b[M]$, Eq. 1 can be simplified :

$$\frac{dQ}{dn_L}(Z) = \frac{1}{2}\Delta H_b \left(1 + \frac{n - Z - r}{\sqrt{(n + Z + r)^2 - 4Zn}}\right)$$

(2)

The function depicted in Eq. 2 exhibits a sigmoidal decrease of the exchanged heat as a function of Z. The larger the r-parameter, the steeper the decrease of $dQ/dn_L$. Note that for Z = n, the function has an inflexion point. In this work, Eq. 2 was utilized to adjust the experimental data, yielding values for the three fitting parameters, $\Delta H_b$, $K_b$ and n. From the determination of binding constant ($K_b$), reaction stoichiometry (n), enthalpy ($\Delta H_b$), the total free energy $\Delta G = -RT\ln K_b$ and entropy $\Delta S = (\Delta H_b - \Delta G)/T$ could be derived, providing a complete thermodynamic profile of the electrostatic complexation reaction.

# 3 – Materials and Methods
## 3.1 – Polymers and surfactants
In the present survey, we report on the anionic-neutral block copolymer poly(acrylic acid)-*b*-poly(acrylamide) and on the oppositely charged surfactant, dodecyltrimethylammonium bromide (DTAB) [6, 33-35]. The synthesis of the copolymer was based on the Madix technology which uses the xanthate as chain-transfer agent in the controlled radical polymerization [37]. Static and dynamic light scattering experiments were performed on polymer solutions to ascertain the weight-average molecular weight $M_W$ and mean hydrodynamic diameter $D_H$ of the chains [35].

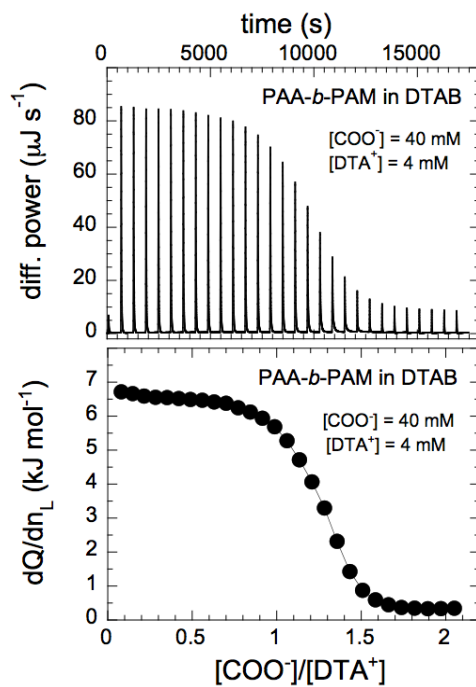

*Figure 2 :* Thermogram (a) and binding isotherm (b) showing the addition of 40 mM COO⁻ (syringe) into 4 mM DTAB solution (cell). The 40 mM COO⁻ corresponded to a weight concentration $c_P = 1.92$ wt. %. In this assay, 10 µL of PANa$_{6.5K}$-b-PAM$_{37K}$ solution was injected every 10 min into the adiabatic cell (Type II experiment). The data points in b) resulted from the integration of the ITC peaks that appear in the upper panel (a).

The molecular weights targeted by the synthesis were 5000 and 30000 g mol⁻¹ for the charged and neutral blocks respectively. The molecular weight of the whole chain as determined from light scattering was slightly higher, at $M_W^P = 43\,500 \pm 1000$ g mol⁻¹. Dynamic light scattering revealed an hydrodynamic diameter $D_H = 11$ nm and size exclusion chromatography a polydispersity index $M_w/M_n = 1.6$. Poly(acrylic acid) is a weak polyelectrolyte and its ionization state depends on the pH. In order to derive the molecular weight of the acrylic acid block, titration experiments were performed. By slow addition of sodium hydroxide 0.1 M, the pH of PANa-b-PAM solutions was varied systematically from acidic to basic conditions. Titration curves and equivalences on copolymers were compared to those obtained on a PANa homopolyelectrolytes with known molecular weight. This procedure allowed us to get for the anionic block a degree of polymerization of 90, yielding $M_W = 6500$ and 37000 g mol⁻¹ for the anionic and neutral block respectively. The degree of polymerization of the charged segment was determined for the acidic form of the polymer. However, the complexation was carried out at neutral pH, *i.e.* in a state where the condensed counterions were sodium ions. In the sequel of the paper, the polymer will be abbreviated as PANa$_{6.5K}$-b-PAM$_{37K}$ [38]. DTAB was purchased from Sigma and used without further purification. The surfactant belongs to the alkyltrimetylammonium class and exhibit an hexagonal mesophase at high concentrations [39]. The critical micellar concentration (cmc) in $H_2O$ is 0.46 wt. % (15 mM) for DTAB [40].

## 3.2 – Mixing protocols

Mixed solutions of surfactant and polymer were prepared according to two different methods. The first method, called *direct mixing* utilized stock surfactant and polymer solutions prepared at the same weight concentration (c = $10^{-2}$ wt. % - 20 wt. %) and same pH (pH 8). The relative amount of each component was monitored by the charge ratio Z = [DTA$^+$]/[COO$^-$]. Z = 1 describes the isoelectric solution characterized by the same number densities of positive and negative chargeable ions. In the mixed solution, the surfactant and polymers concentrations were calculated according to $c_{Surf} = cZ(z_0+Z)^{-1}$ and $c_{Pol} = cz_0(z_0+Z)^{-1}$ with $z_0 =$ 1.567. Once prepared by direct mixing, the samples were allowed to rest for a week before being studied, yielding fully reproducible results.

In the second protocol, surfactants were titrated by step addition of polymers. The addition was performed by injecting 10 µl of the polymer solution every 10 minutes, each injection lasting 20 seconds. Experiments were performed over a periods of 4 – 6 hours, with a range in charge ratio covered being typically between Z = 0 and Z = 2. Another difference with direct mixing is that the total concentration in active matter increased during titration. In this work, titration was carried out either by adding surfactant to polymer, or the reverse, by adding polymer to surfactant. These experiments will be described as Type I and Type II respectively. As demonstrated in this paper, these two processes were found to provide different results. As shown recently using light scattering measurements in the very dilute regime, the surfactant-based complexes DTAB/PANa$_{6.5K}$-*b*-PAM$_{37K}$ exhibited a critical association concentration (cac) estimated at $c_{cac}$ = 1.1×10$^{-2}$ wt. %, corresponding to a DTAB concentration of 4.3×10$^{-3}$ wt. % or 0.14 mM [38].

### 3.3 – Isothermal titration calorimetry

Isothermal titration calorimetry (ITC) was performed using a Microcal VP-ITC calorimeter (Northampton, MA) with the normal cell (1.4643 mL) at 25 °C (Fig. 1). In this work, either the DTAB was injected into a PANa$_{6.5K}$-*b*-PAM$_{37K}$ solution or PANa$_{6.5K}$-*b*-PAM$_{37K}$ was injected into the DTAB solution. The concentrations were expressed in molar concentrations of charges, noted [DTA$^+$] for the surfactant and [COO$^-$] for the polymer. A 1 mM [COO$^-$] solution corresponded to $c_{Pol}$ = 0.048 wt. %. Each titration consisted of a preliminary 2 µL injection followed by 28 subsequent 10 µL injections at 10 - 20 min intervals. The syringe is tailor-made such that the tip acts as a blade-type stirrer to ensure an optimum mixing efficiency at 307 rpm. A typical ITC experiment including the thermogram and binding isotherm is shown in Fig. 2a and 2b respectively. There, a 40 mM COO$^-$ solution was injected through the syringe into a 4 mM DTAB solution. Control experiments were carried out for both surfactant and polymer to determine the heats of ligand dilution. The thermal responses of the ligand dilution were then subtracted to obtain the heat of binding.

### 3.4 – Light scattering

Static and dynamic light scattering were performed on a Brookhaven spectrometer (BI-9000AT autocorrelator, λ = 632 nm) for measurements of the Rayleigh ratio $\mathcal{R}(q,c)$ and of the collective diffusion constant D(**c**). The Rayleigh ratio was obtained from the scattered intensity I(q,c) measured at the wave-vector q according to $\mathcal{R}(q,c) = \mathcal{R}_{std}(n_0/n_{Tol})^2(I(q,c) - I_S)/I_{Tol}$, where $\mathcal{R}_{std}$ and $n_{Tol}$ are the standard Rayleigh ratio and refractive index of toluene, $I_S$ and $I_{Tol}$ the intensities measured for the solvent and for the toluene in the same scattering configuration and $q = \frac{4\pi n_0}{\lambda}\sin(\theta/2)$ (with $n_0$ the refractive index of the solution and $\theta$ the scattering angle). For the determination of the molecular weight of the polymer, the refractive index was measured on a Chromatix KMX-16 differential refractometer at room temperature. $dn_0/dc$-values for DTAB and PANa$_{6.5K}$-$b$-PAM$_{37K}$ solutions were found at 0.231 cm$^3$ g$^{-1}$ and 0.157 cm$^3$ g$^{-1}$ respectively [38]. In dynamical mode, the collective diffusion coefficient $D(c)$ was measured in the range $c = 0.01$ wt. % – 1 wt. %. From the value of $D(c)$ extrapolated at $c = 0$, the hydrodynamic radius of the colloids was calculated according to the Stokes-Einstein relation, $D_H = k_B T/3\pi\eta_S D_0$, where $k_B$ is the Boltzmann constant, T the temperature (= 298 K) and $\eta_S$ (= $0.89\times10^{-3}$ Pa s) the solvent viscosity. The autocorrelation functions of the scattered light were interpreted using both the method of cumulants and the CONTIN fitting procedure provided by the instrument software.

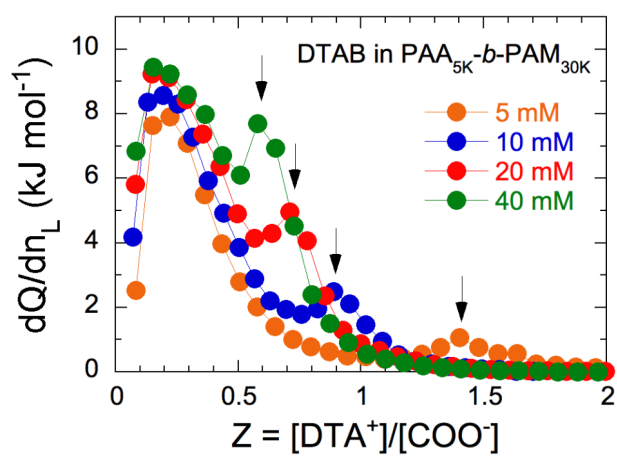

*Figure 3 :* *Binding isotherms resulting from the addition of a DTAB solution to a PANa$_{6.5K}$-b-PAM$_{37K}$ solution at pH 8 and T = 25°C (Type I experiment). The DTAB concentrations were 5, 10, 20 and 40 mM, whereas those of the polymers corresponded to [COO$^-$] = 0.5, 1, 2 and 4 mM respectively. The enthalpies of binding were positive, indicating an endothermic reaction. In each data set, a secondary ITC peak occurred at the critical charge ratio $Z_C$ (shown by arrows).*

## 4 – Results
### 4.1 - Isothermal titration calorimetry (ITC)

ITC experiments were performed in two ways [26]. Either the surfactant solution was injected into the polymer solution (called Type I experiment) or the polymer solution was injected into the surfactant one (Type II experiment). With the notations of the theory Section, the results of Type I runs were analyzed assuming PANa$_{6.5K}$-$b$-PAM$_{37K}$ as the macromolecule and DTAB as the ligand. For runs of Type II, the surfactant was playing the role of the macromolecule and the polymer that of the ligand. As shown below, the mixing order is a crucial factor in the electrostatic complexation process.

Fig. 3 shows the binding isotherms resulting from the addition of DTAB solution to PANa$_{6.5K}$-$b$-PAM$_{37K}$ solution at different concentrations, pH 8 and T = 25°C (Type I experiment). The DTAB concentrations were comprised between 5 and 40 mM, whereas those of the charges borne by the polymers were 10 times lower, corresponding to [COO$^-$] = 0.5 to 4 mM. The surfactant concentrations were chosen so as to explore different cases, namely cases where the surfactants were injected under the form of micelles (c$_{Surf}$ > c$_{cmc}$ = 15 mM) and cases where the surfactants were dispersed as unimers (c$_{Surf}$ < c$_{cmc}$). The derivative of the heat changes was expressed as a function of the charge ratio Z. Despite the differences in aggregation state, the four titration curves of Fig. 3 exhibited the typical sigmoidal decrease of binding isotherms as more ligands were injected. In Fig. 3, the enthalpy associated with the DTAB/PANa$_{6.5K}$-*b*-PAM$_{37K}$ binding is positive, indicating that the process is endothermic [23, 24]. The thermal responses displayed in addition a second process at the critical charge ratio $Z_C$ (arrows) which showed up as a secondary endothermic peak. The sigmoidal decrease at low Z was sometimes described as a primary peak [23, 24]. The secondary ITC peak was found to be concentration dependent, varying from $Z_C$ = 1.42 at [DTA$^+$] = 5 mM to $Z_C$ = 0.60 at [DTA$^+$] = 40 mM.

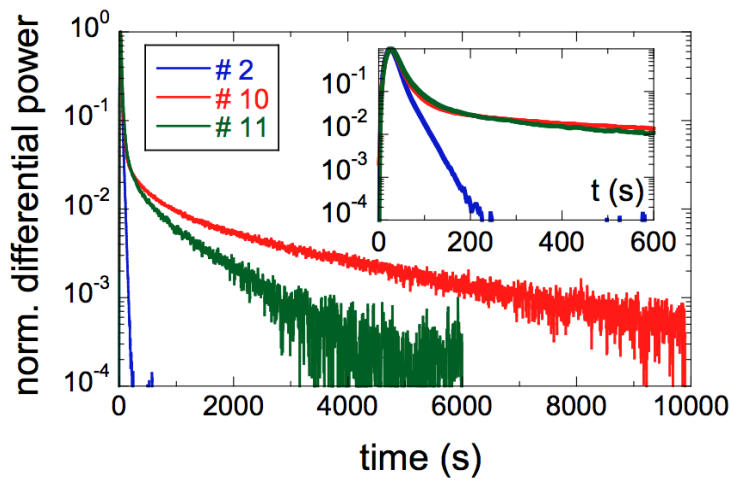

*Figure 4 :* *Normalized differential power plotted as a function of the time for different injections. The experiments was of Type I, using [DTA$^+$] = 20 mM, [COO$^-$] = 2 mM and a time interval between injections of 180 min. The thermogram for the second injection was representative of the response function of the calorimeter. For the injections located around $Z_C$ (here the 10$^{th}$ and 11$^{th}$ injections), the return towards equilibrium exhibited a single exponential decay, with characteristic time τ=3200 s and 1020 s respectively.*

In the range of the secondary peak, the differential power delivered by the calorimeter to keep the cell temperature constant exhibited a singular behavior, namely a transient response that lasted up to several hours after a 10 μl-injection. Fig. 4 illustrates the slow kinetics associated with the secondary peak. In the main frame, the normalized differential power was plotted as a function of the time for different injections, in an experiment carried out at [DTA$^+$] = 20 mM and [COO$^-$] = 2 mM. In contrast to the data in Fig. 3, the interval between two injections was fixed at 180 min. The inset illustrates that the response following the second injection (Z << $Z_C$) was rapid and few minutes were necessary for the differential power to return to its baseline. This profile was representative of the response function of the calorimeter. For the injections located around $Z_C$ (here the 10$^{th}$ and 11$^{th}$ injections), the return towards equilibrium took up to several hours, indicating that the secondary endothermic peak was linked to a very slow process. Interestingly, all long-time transients could be fitted with a single exponential decay, with a unique characteristic time τ. For the second injection, one had τ = 22 s, whereas for the 10$^{th}$ and 11$^{th}$ injection, τ was 3200 s and 1020 s respectively. Note that this slow kinetic was observed for all secondary peaks observed in Fig. 3.

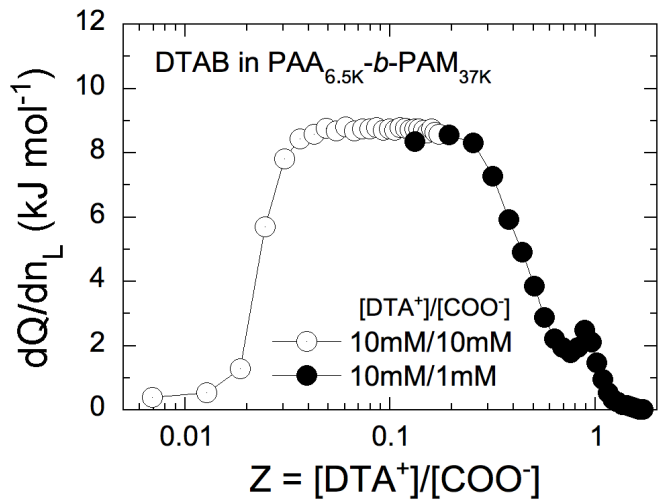

*Figure 5 :* *Type I ITC experiment carried out by titrating PANa$_{6.5K}$-b-PAM$_{37K}$ solutions at [COO$^-$] = 1 mM (open symbols) and 10 mM (close symbols) by a 10 mM DTAB solution. Between Z = 0.02 and Z = 0.2, the rate of heat exchange exhibits a plateau around dQ/dn$_L$ = + 10.7 kJ mol$^{-1}$.*

In Fig. 5, Type I ITC runs were carried out by titrating PANa$_{6.5K}$-b-PAM$_{37K}$ solutions at [COO$^-$] = 1 mM (open symbols) and 10 mM (close symbols) of charges by a 10 mM solution of DTAB. Doing so, we could expand the range of charge ratios as compared to Fig. 3 (Z = 5×10$^{-3}$ to Z = 2). At Z = 0.02, the rate of exchanged heat increased rapidly and reached a saturation plateau around dQ/dn$_L$ = + 10.7 kJ mol$^{-1}$, in good agreement with the value determined at lower c$_{Pol}$. The isotherms in Figs. 3 and 5 exhibit strong similarities with data reported by Wang and Tam [23, 24] on DTAB/PANa or by Zhu and Evans [18] on plasmid DNA titrated by cationic surfactants, or by multivalent counterions such as cobalt hexammine and spermidine [27]. In particular, the existence of the secondary endothermic peak at relatively high added surfactant (e.g. beyond the reaction stoichiometry in some cases) was already disclosed by these authors [17, 18, 22, 23].

Type II experiments were carried out by considering the surfactants as the macromolecules and the polymers as the ligands. Fig. 6 shows the heat detected upon gradual injection of PANa$_{6.5K}$-b-PAM$_{37K}$ at [COO$^-$] molar concentrations 10 mM to 40 mM into a DTAB solution as a function of the inverse charge ratio 1/Z. As in Fig. 3, the initial surfactant concentrations in the measuring cell were adjusted to be 1/10 of that of the [COO$^-$]. For the three concentrations tested, the thermograms in Fig. 6 looked similar. They displayed an initial positive plateau around dQ/dn$_L$ = + 7 kJ mol$^{-1}$ followed by a sigmoidal decrease as titration proceeded. The enthalpy associated with the DTAB/PANa$_{6.5K}$-b-PAM$_{37K}$ binding was positive, indicating again an endothermic process. The magnitude of the heat exchanged was however lower that that found in type I experiments. Note that the inflexion points of the thermal responses were located at inverse charge ratio larger than 1, and that the secondary endothermic peaks seen at high surfactant concentrations did not show up.

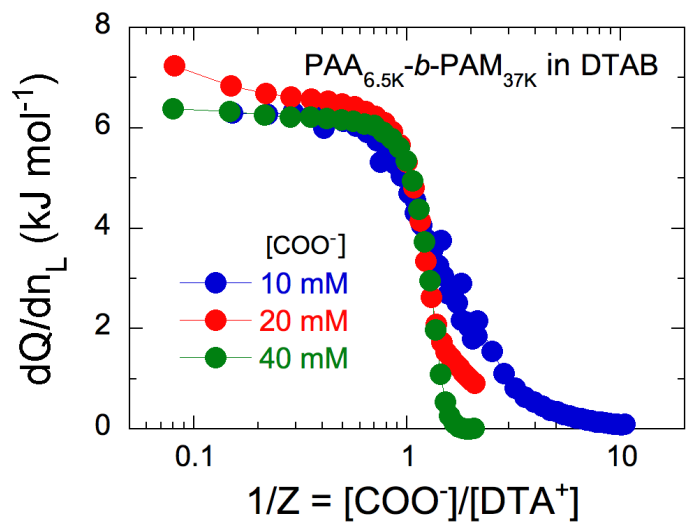

***Figure 6 :*** *Binding isotherms resulting from the injection of PANa$_{6.5K}$-b-PAM$_{37K}$ solutions into DTAB solutions at pH 8 and T = 25°C (Type II experiment). The carboxylate concentrations corresponded to [COO$^-$] = 10, 20 and 40 mM, whereas those of the surfactant were [DTAB] = 1, 2 and 4 mM respectively. The enthalpies of binding were positive, indicating an endothermic reaction.*

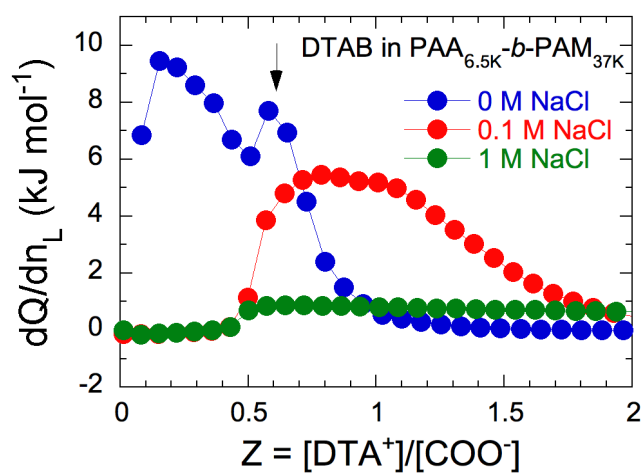

***Figure 7 :*** *Type I binding isotherms resulting from the addition of a 40 mM DTAB solution to a PANa$_{6.5K}$-b-PAM$_{37K}$ solution at [COO$^-$] = 4 mM at different ionic strengths [NaCl] = 0 M, 0.1 M and 1 M. The experimental configuration was that of Fig. 3. The binding enthalpy decreases with increasing added salt, proving the electrostatic nature of the association.*

To prove that the processes taking place in the titration experiments were driven by electrostatics, Type I ITC runs were conducted at increasing sodium chloride concentrations. Data were collected at [DTAB] = 40 mM and [COO$^-$] = 4 mM. Fig. 7 compares the binding isotherms at [NaCl] = 0.1 M and 1 M to that obtained without added salt. The data in Fig. 7 show that the magnitude of the ITC signal varied inversely with the salt concentration, and so did the binding enthalpy. At [NaCl] = 1 M, the thermal response was comparable to that of the dilution of the polymers in a 1 M brine, indicating that at this ionic strength DTAB and PAA$_{6.5K}$-*b*-PAM$_{37K}$ did not interact [41]. In Fig. 7, the vanishing of the secondary ITC peak at high ionic strength was also noticed.

## 4.2 - Light scattering

Fig. 8 and 9 display the correlations between the thermodynamics and structure of the co-assembly in Type I and Type II experiments, respectively. For these experiments, the molar concentrations for the charged species were [DTAB] = 20 mM and [COO$^-$] = 2 mM. In Fig. 8, the upper panel shows a double scale representation of the Rayleigh ratio $\mathcal{R}_\theta(q,c)$ (left) at q = 1.87×10$^{-3}$ Å$^{-1}$ and of the hydrodynamic diameter $D_H$ (right) versus Z. The static and dynamic light scattering experiments were performed in the same conditions as in ITC, that is by titrating the polymer solution with that of the surfactant. Each data point in the figure corresponds to the addition of 10 µl of DTAB solution. The time interval between two injections was also identical to that of the ITC assays. At low Z, typically below 0.20, the scattering intensity remained at the level of the pure polymer. In this range, $D_H$ could not be measured because of a weak scattering signal. In a second regime, 0.20 < Z < 0.65, the Rayleigh ratio increased linearly and leveled off for Z > 1. In this regime, dynamic light scattering showed the existence of few but large aggregates with hydrodynamic diameters $D_H$ ~ 200 – 300 nm.

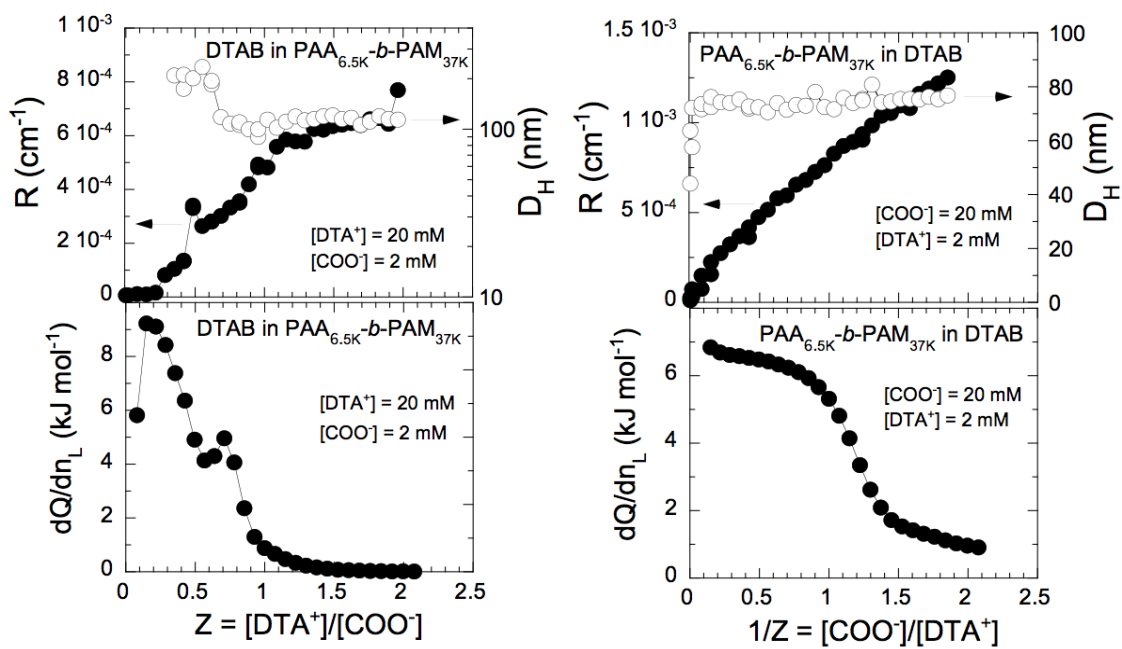

*Figure 8 :* Upper panel : Rayleigh ratio $\mathcal{R}_\theta$ (left scale) at measured at scattering angle θ = 90° and hydrodynamic diameter $D_H$ (right scale) plotted versus Z. The light scattering experiment was performed in the same conditions as the ITC shown in the lower panel. Lower panel : Binding isotherms found from the addition of a 20 mM DTAB solution to a PANa$_{6.5K}$-b-PAM$_{37K}$ solution at [COO$^-$] = 2 mM (data from Fig. 3)
*Figure 9 :* Same as in Fig. 8 for Type II experiments.

With increasing Z, the size diminished and stabilized above $Z_C$ = 0.65 at $D_H$ = 110 ± 10 nm. In this range, the analysis of the autocorrelation functions revealed a single exponential decay in agreement with monodisperse colloids [1, 42]. The light scattering data of Fig. 8 are in good agreement with those reported on this surfactant/polyelectrolyte system [33, 43]. As compared to the ITC data (lower panel in Fig. 8), a good correlation between the $D_H$-plateau above Z =

0.65 and the secondary ITC peak was observed too. In Type II experiments (polymer in surfactant), light scattering revealed different features. The upper panel of Fig. 9 shows that as copolymers were injected into a 2 mM DTAB solution, the Rayleigh ratio increased steadily, whereas the hydrodynamic diameter exhibited a sudden jump already at the first injection. This jump was followed by a plateau at $D_H = 75 \pm 5$ nm over the whole range. The main differences with Type I assays were that even at very low amount of added charges, large colloidal complexes formed and that the sizes of these aggregates were smaller (75 nm *versus* 110 nm). Such a result illustrates the fact that the way of mixing oppositely charged systems is crucial [33, 44, 45]. As for the Rayleigh ratio, the calorimetric titration of the cationic surfactants by the anionic copolymers (lower panel) appeared as a continuous and regular process and each injection yielded the formation of new aggregates with identical size.

# 5 - Discussion
## 5.1 – Interpretation of the ITC secondary peak
Fig. 10 shows the critical charge ratio $Z_C$ as a function of the total concentration c. The ITC-data were determined from the position of the secondary peak showing up in the titration isotherms of Fig. 3. As already mentioned, $Z_C$ decreased with increasing concentration. Also displayed in Fig. 10 are the critical values found from light scattering experiments [33]. In Ref. [33] from where the light scattering data are stemming, the solutions were prepared by *direct mixing*, from stock solutions made at the same concentration and pH. The mixing was controlled by pouring rapidly the surfactant in the polymer solution at the desired charge ratio. The specimens were allowed to rest at room temperature for a week before being investigated by light scattering. As a result, the Rayleigh ratios plotted as a function of Z exhibited a steep jump at $Z = Z_C$ (e.g. see Fig. 2 in Ref. [33]). The results were obtained for c = 0.1 wt. % to 20 wt. % and showed that $Z_C$ decreases with increasing c. The behavior observed on samples prepared by *direct mixing* was slightly different from that made by *titration*. In Fig. 8 for instance, the Rayleigh ratio grows progressively and flattens out above the charge stoichiometry, but it does not display the steep increase of Ref. [33]. The slow transients found in the thermograms around $Z_C$ (Fig. 4) could explain these differences. The main result of Fig. 10 remains the good agreement between the ITC and light scattering experiments. Over three decades in concentrations, $Z_C$ was found to decrease as a power law : $Z_C(c) = 0.012 \times c^{-0.6}$ (continuous line). These results have a direct implication. They allow us to ascribe the ITC secondary peaks of Fig. 3 to the formation of hierarchical core-shell complexes. Similar conclusions were drawn from the thermodynamics of binding and condensation between DNA and oppositely charged surfactants [18] or multivalent counterions [27].

## 5.2 – Fitting the binding isotherms
The Multiple Non-interacting Sites model was used to adjust the results of the binding isotherms of Figs. 3, 6 and 7. In Type I experiments, Eq. 2 was applied and allowed the determination of the binding enthalpy ($\Delta H_b$), the binding constant ($K_b$) and the reaction stoichiometry n. According to the MNIS model, n is the value of the charge ratio $[DTA^+]/[COO^-]$ at the inflexion point and stands for the number of surfactant molecules per carboxylate monomer that were involved in the complexation mechanism. In type II assays, the same expression was considered but in order to take into account the conventions adopted in the paper, Z was replaced by Z' = 1/Z and n by n' = 1/n. The other adjusting parameters remained unchanged. The continuous lines in Figs. 11a and 11b result from best fit calculations using Eq. 2 for the three concentration pairs [L]/[M] = 10mM/1mM, 20mM/2mM and 40mM/4mM. For the type I experiments, the data points of the ITC secondary peaks were not taken into account [17].

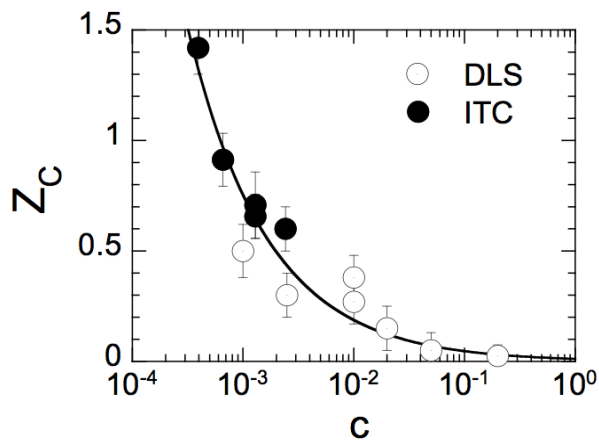

*Figure 10 :* *Critical charge ratio $Z_C$ as a function of the total concentration c as determined from isothermal titration calorimetry (closed symbols) and from static light scattering (open symbols)* [33]. *The continuous line resulted from best fit calculations using the power law : $Z_C(c) = 0.012 \times c^{-0.6}$.*

The fitting parameters for type I and II titration experiments are listed in Table I and II, respectively. For type I experiments, the binding enthalpy and the stoichiometry remained concentration independent at $\Delta H_H^I = +10.65 \pm 0.15$ kJ mol$^{-1}$ and $n^I = 0.48 \pm 0.07$. As already mentioned, positive binding enthalpies indicate an endothermic processes. Values of the stoichiometry ratio $n^I$ below 1 also suggest that the complexation occurred with an excess of anionic charges. More precisely, for DTAB micelles with an aggregation number $53 \pm 3$ [35], a stoichiometry of $0.48 \pm 0.07$ corresponds to a number of carboxylates involved in the process of $110 \pm 20$. This result suggests that for the Type I mode the complexation results in the association of about one block copolymer per surfactant micelle. We recall that the degree of polymerization of the charged block in PANa$_{6.5K}$-b-PAM$_{37K}$ is 90. In such a process, only half of the anionic charges of the polyelectrolyte block are complexed onto the surface of the micelles. In Table I, the binding constant $K_b^I$ was found to decrease with increasing concentration as $K_b^I \sim c^{-2}$, a result that is not included the MNIS model. The concentration dependence for the binding constant had another effect, namely to induce some weak variation in the free energy and entropy $\Delta G^I$ and $\Delta S^I$ *versus* concentration. Values for $\Delta G^I \sim -10$ kJ mol$^{-1}$ and $\Delta S^I = 70$ kJ mol$^{-1}$ K$^{-1}$ are however in excellent agreement with those derived by Wang and Tam from the titration of poly(acrylic acid) homopolyelectrolytes by DTAB. For type II experiments, the binding enthalpy was lower than in type I experiments by 30 % ($\Delta H_H^{II} = 7.2 \pm 1$ kJ mol$^{-1}$), whereas the stoichiometry was slightly higher, at $n^{II} = 0.75 \pm 0.05$. Assuming again that DTAB micelles have an aggregation number of 53, a reaction stoichiometry of $n^{II} = 0.75$ corresponds to a number of carboxylates involved in the process of $70 \pm 15$. More surprising for this data set was the increase of the binding constant $K_b^{II}$ with increasing concentration (Table II). The discrepancies between the reaction stoichiometry in Type I and II titration are discussed in the next section.

| Type I Surfactant in Polymer | $\Delta H_b^I$ kJ mol$^{-1}$ | $K_b^I$ M$^{-1}$ | $n^I$ | $\Delta G^I$ kJ mol$^{-1}$ | $\Delta S^I$ J mol$^{-1}$K$^{-1}$ |
|---|---|---|---|---|---|
| 5 mM in 0.5 mM | +10.8 | 4.8×10$^4$ | 0.41 | -11.6 | +75.2 |
| 10 mM in 1 mM | +10.7 | 2.3×10$^4$ | 0.46 | -10.8 | +72.3 |
| 20 mM in 2 mM | +10.6 | 1.5×10$^4$ | 0.51 | -10.4 | +70.2 |
| 40 mM in 4 mM | +10.5 | 7.4×10$^3$ | 0.55 | -9.6 | +67.4 |

**Table I**: List of the fitting parameters $\Delta H_b^I$, $K_b^I$ and $n^I$ derived for Type I titrations. $\Delta H_b^I$, $K_b^I$ and $n^I$ denote the binding enthalpy, the binding constant and the reaction stoichiometry, respectively and allowed us to derive the variations in free energy $\Delta G^I$ and entropy $\Delta S^I$ in Type I reactions. Positive binding enthalpies indicate endothermic processes.

| Type II Polymer in Surfactant | $\Delta H_b^{II}$ kJ mol$^{-1}$ | $K_b^{II}$ M$^{-1}$ | $n^{II}$ | $\Delta G^{II}$ kJ mol$^{-1}$ | $\Delta S^{II}$ J mol$^{-1}$K$^{-1}$ |
|---|---|---|---|---|---|
| 10 mM in 1 mM | +8.3 | 3.1×10$^3$ | 0.67 | -8.7 | +56.9 |
| 20 mM in 2 mM | +7.2 | 1.1×10$^4$ | 0.79 | -10.0 | +57.5 |
| 40 mM in 4 mM | +6.3 | 2.5×10$^4$ | 0.79 | -10.9 | +57.5 |

**Table II**: Same as Table I for Type II experiments.

### 5.3 – Interpretation of the ITC results

The fact that the thermodynamic parameters acquired by the MNIS model are different in type I and II experiments indicates that the order of mixing oppositely charged species is an important parameter. The results of Fig. 8 and 9 clearly demonstrate that not only the microstructures formed are different, but also the thermodynamics. Fig. 12 illustrates the scenario of the complexation in Type I titration. In this diagram, three regimes can be distinguished. At very low charge ratios (Z < 0.01), the amount of exchanged heat is negligible. This first regime is characterized by a non-cooperative binding of the surfactants on the polymers [20], as expected from concentrations below the cac. The value of the cac was determined independently for DTAB-PANa$_{6.5K}$-b-PAM$_{37K}$ at $c_{cac}$ = 1.1×10$^{-2}$ wt. % [38]. Assuming that in the experiment of Fig. 12 surfactants and polymers interact at the reaction stoichiometry ($n^I$ = 0.48), the value of the cac corresponds to Z = 0.02, in good agreement with the steep increase of the heat exchange rate.

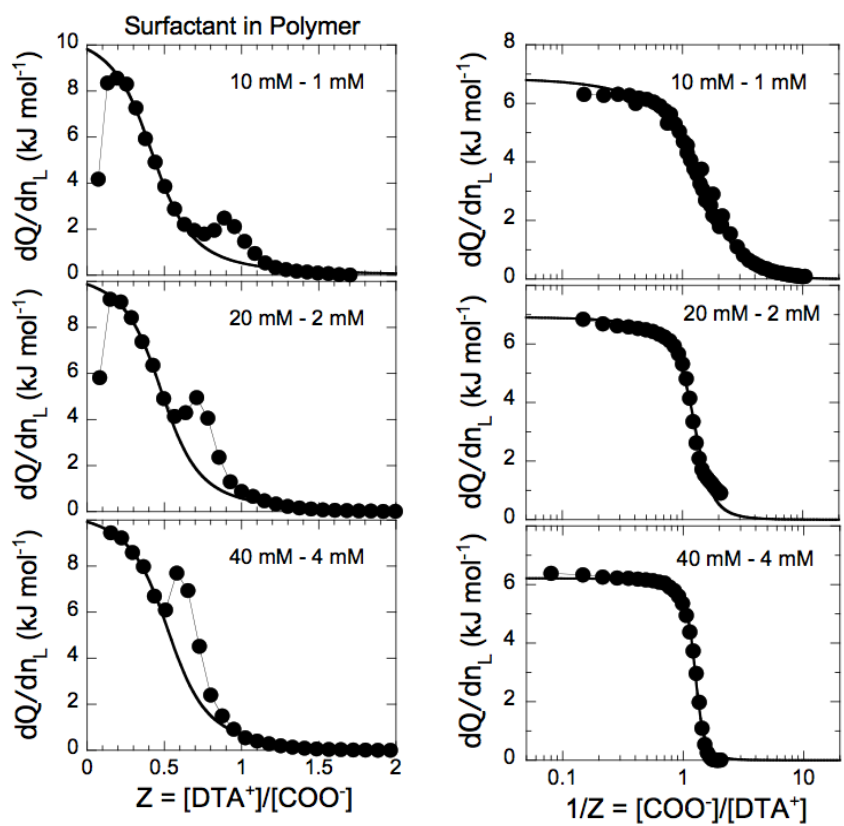

*Figure 11 :* a) Binding isotherms of Type I experiments, together with best fit calculations using the Multiple Non-interacting Sites model (Eq. 2). The enthalpy of binding ($\Delta H_b^I$), the binding constant ($K_b^I$) and the reaction stoichiometry $n^I$ used as adjusting parameters are listed in Table I. The data points of the secondary peak were not taken into account for the fitting. b) Same as in Fig. 11a for Type II experiments. The fitting parameters $\Delta H_b^{II}$, $K_b^{II}$ and $n^{II}$ are provided in Table II.

In a second regime, the total concentration of interacting species is above the cac and $dQ/dn_L$ exhibits a plateau. In this range, the binding is cooperative and individual micelles made from surfactants and from the polyelectrolyte block form spontaneously. Neutron experiments performed on DTAB/PANa$_{6.5K}$-*b*-PAM$_{37K}$ established conclusively the existence of single DTAB micelles below the critical charge ratio $Z_C$ [33]. Interestingly, the formation of micelles resulting from an electrostatic complexation process is endothermic [10, 16, 17, 20, 22-26], whereas the micellization of surfactants (*i.e.* in absence of polymers) was shown to be strongly exothermic [46, 47]. These results and interpretation are in excellent agreement with the observations made on polyelectrolytes [20] or on DNA [18] complexed with oppositely charged surfactants. In Fig. 12, the decrease of the thermal response for $Z > 0.2$ is then related to the titration of the anionic charges. It is accompanied by the formation of individual micelles, and not with that of colloidal complexes. For this process, the stoichiometry is 0.48 and corresponds approximately to one polymer per micelle. The formation of core-shell complexes of larger sizes ($D_H \sim 100$ nm) occurred at higher charge ratios, above $Z_C$. This was demonstrated in Fig. 10 which shows a very good correlation between the Z's of the complex formation and the secondary ITC peak. This later phenomenon is strongly concentration dependent, whereas the polyelectrolyte induced micelle formation is not. In Type II experiments, hierarchical clusters of micelles developed even at the first injection of polymer solution, corresponding to Z = 200. This result was confirmed by light scattering in Fig. 9.

There, the concentration of reacting polymers and surfactants (assuming a stoichiometry of $n^{II}$ = 0.75) is of the order of the cac. In Type II mode, hierarchical core-shell aggregates form spontaneously without passing by the intermediate state of single micelles decorated by one copolymer. In a diagram that would be equivalent to that of Fig. 12, there would be only a unique regime in type II experiments : colloidal complexes are present for at all Z's probed by ITC. Calorimetry experiments at very high Z (Z > 1000) remain difficult to perform because of the lack of sensitivity at such low added polymer concentrations. The values of the reaction stoichiometries in type I and II experiments deserves also a comment. Their differences being larger than the experimental uncertainties ($n^I$ = 0.48 ± 0.07 and $n^{II}$ = 0.75 ± 0.05), it suggests that the formation of individual micelles and that of hierarchical clusters occur at different polymer-to-surfactant ratio. In type I (resp. II) experiments, we have found approximately 1 (resp. 0.7) polymer per micelle. According to this scheme, the secondary ITC peak can now be interpreted as a rearrangement of single micelles into large clusters, this rearrangement being associated with a change of stoichiometry, $n^I \to n^{II}$ and with very slow kinetics.

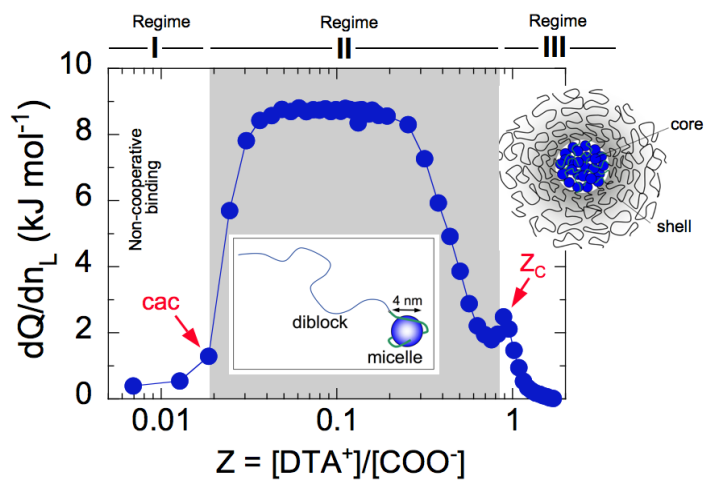

*Figure 12 :* Illustration of the complexation scenario in Type I titration. Regime I is characterized by a non-cooperative binding of the surfactants on the polymers. In Regime II, above the cac, the binding is cooperative and individual micelles made from surfactants and from the polyelectrolyte block form spontaneously. Regime III corresponds to the formation of core-shell complexes of large sizes, $D_H \sim 100$ nm.

## 6 - Conclusion

In the present paper, we investigate the interactions between oppositely charged surfactants and double-hydrophilic copolymers using a combination of isothermal titration calorimetry and light scattering. ITC experiments were performed by titrating either the copolymers by the surfactants, or the surfactants by the copolymers. With Type I assays, two important results previously reported with homopolyelectrolytes were first confirmed : *i)* in the concentration range investigated, the binding isotherms did not depend on the aggregation state of the surfactant; *ii)* The reaction was endothermic, indicating a co-assembly mechanism dominated by the entropy of the released counterions [25, 27, 31, 48]. In addition to this, we observed that the binding isotherms corresponding to different mixing orders yielded different results for the binding enthalpy and reaction stoichiometry. For Type I experiments, the co-assembly occurred in a two-step process. Below a critical charge ratio $Z_C$, there is first the formation of single surfactant micelles, these micelles being decorated by a unique polymer. Above $Z_C$, these single micelles rearrange themselves into large colloidal core-shell complexes. The signatures of this transformation are the presence of a secondary endothermic peak at $Z_C$ and a very slow relaxations of the differential power applied by the calorimeter to reach thermal equilibrium. It is important to recall here that in a previous paper, we determined the charge stoichiometry for DTAB/PANa$_{6.5K}$-*b*-PAM$_{37K}$ core-shell colloids. Based on a structural study of the core-shell assembly by neutron scattering, we found n = 0.66 ± 0.06. This later value is in excellent agreement with the present value of 0.75 ± 0.05. With these additional results, we also confirm the result that colloidal complexes form with a excess of charges coming from the copolymers. For Type II experiments, the phase diagram describing the co-assembly is composed by a unique regime, namely that of core-shell complexes present at all charge ratios. Since many oppositely charged species exhibit the same thermodynamic features than those discussed here [17, 23, 27], the above conclusions might certainly apply to these systems.

**Acknowledgement**
We thank Laurent Bouteiller, Jean-Paul Chapel, Olivier Sandre, Régine Perzynski for numerous and fruitful discussions during the course of this work. We are also grateful to Watson Loh for his critical readings and comments on the manuscript. This research was supported in part by Rhodia (France), by the Agence Nationale de la Recherche under the contract BLAN07-3_206866, by the European Community through the project : "NANO3T—Biofunctionalized Metal and Magnetic Nanoparticles for Targeted Tumor Therapy", project number 214137 (FP7-NMP-2007-SMALL-1) and by the Région Ile-de-France in the DIM framework related to Health, Environment and Toxicology (SEnT).